\documentclass[prd,twocolumn,showpacs,preprintnumbers,amsmath,amssymb,superscriptaddress,floatfix,nofootinbib]{revtex4-2}
\usepackage{graphicx}
\usepackage{bm}
\usepackage{amsmath}  
\usepackage{amsfonts}
\usepackage{amssymb}
\usepackage{color}
\usepackage{multirow}
\usepackage{booktabs} 
\usepackage{overpic} 
\usepackage[colorlinks, citecolor=blue,anchorcolor=red,menucolor=red, linkcolor=red,filecolor=red,runcolor=red,urlcolor=blue,frenchlinks=red]{hyperref}
\usepackage{xeCJK} 

\begin{document}

\title{
Dynamical resonances $\Lambda(1405)$, $\Lambda(1670)$, and $f_0(980)$ in the process $\Lambda_c^+ \to p K^+ K^-$}
 
\author{Sheng-Chao Zhang}
\affiliation{School of Physics, Zhengzhou University, Zhengzhou 450001, China}

\author{Gui-Ye Guo}
\affiliation{School of Physics and Electronics, Henan University, Kaifeng 475004, China}

\author{Guan-Ying Wang}~\email{wangguanying@henu.edu.cn}
\affiliation{School of Physics and Electronics, Henan University, Kaifeng 475004, China}

\author{Li-Sheng Geng}~\email{lisheng.geng@buaa.edu.cn}
 \affiliation{Sino-French Carbon Neutrality Research Center, \'{E}cole Centrale de P\'{e}kin/School
of General Engineering, Beihang University, Beijing 100191, China}
\affiliation{School of Physics, Beihang University, Beijing 102206, China}
\affiliation{Peng Huanwu Collaborative Center for Research and Education, Beihang University, Beijing 100191, China}
\affiliation{Beijing Key Laboratory of Advanced Nuclear Materials and Physics, Beihang University, Beijing 102206, China }
\affiliation{Southern Center for Nuclear-Science Theory (SCNT), Institute of Modern Physics, Chinese Academy of Sciences, Huizhou 516000, China}
\vspace{0.5cm}

\author{En Wang}~\email{wangen@zzu.edu.cn}
\affiliation{School of Physics, Zhengzhou University, Zhengzhou 450001, China}

\author{Ju-Jun Xie}~\email{xiejujun@impcas.ac.cn}
\affiliation{Southern Center for Nuclear-Science Theory (SCNT), Institute of Modern Physics, Chinese Academy of Sciences, Huizhou 516000, China}
\affiliation{State Key Laboratory of Heavy Ion Science and Technology, Institute of Modern Physics, Chinese Academy of Sciences, Lanzhou 730000, China}
\affiliation{School of Nuclear Sciences and Technology, University of Chinese Academy of Sciences, Beijing 101408, China}

\date{\today}

\begin{abstract}
Motivated by new BESIII measurements of the singly Cabibbo-suppressed process $\Lambda_c^+ \to  p K^+ K^- $, we investigate the $\Lambda_c^+ \to  p K^+ K^-$ reaction by taking into account contributions from the low-lying excited baryons $\Lambda(1405)$ and $\Lambda(1670)$, which are dynamically generated via the $S$-wave pseudoscalar meson-octet baryon interaction. In addition, contributions from the light scalar mesons $f_0(980)/a_0(980)$ dynamically generated via the $S$-wave pseudoscalar meson-pseudoscalar meson interaction, as well as the vector meson $\phi(1020)$, are also considered. The theoretical calculations could reproduce the invariant mass distributions of $K^- p$ and $K^+K^-$ measured by the BESIII Collaboration well. Specifically, a near-threshold enhancement and a cusp structure around 1670~MeV in the $pK^-$ invariant mass spectrum are obtained, which are associated with the $\Lambda(1405)$ and $\Lambda(1670)$ resonances, respectively. Furthermore, contributions from the scalar meson $f_0(980)$ and vector meson $\phi(1020)$ are clearly observed in the $K^+ K^-$ invariant mass distribution. Future high-precision measurements by the BESIII, Belle II, and the proposed Super Tau-Charm Facility experiments will substantially deepen our understanding of the intrinsic properties of the low-lying excited $\Lambda$ states and the $f_0(980)/a_0(980)$ meson.

\end{abstract}


\maketitle

\section{Introduction}
Weak decays of charmed baryons provide an ideal platform to simultaneously explore nonperturbative strong dynamics contained within final-state interactions (FSIs)~\cite{Geng:2018upx,Jia:2024pyb} and the fundamental weak interaction described by the Cabibbo-Kobayashi-Maskawa (CKM) matrix~\cite{Miyahara:2015cja}. Among all charmed baryons, the ground-state $\Lambda_c^+$ is the first charmed hadron observed~\cite{Knapp:1976qw}, and its multi-body decays  have attracted mounting experimental and theoretical interest over the past decades, as they offer a unique window for investigating exotic hadron resonances~\cite{Oset:2016lyh,BESIII:2018qyg,BESIII:2023sdr,Belle:2022pwd,LHCb:2019ldj}.  
For instance, roles of the scalar mesons $a_0(980)$ and $f_0(980)$ have been investigated in the decays $\Lambda_c^+ \to pK^+K^-$ and $p\pi^+\pi^-$~\cite{Wang:2020pem}, as well as in $\Lambda_c^+ \to \pi^0\eta p$~\cite{Feng:2020jvp} and $\Lambda_c^+ \to \pi^+\eta n$~\cite{Li:2025gvo,Li:2026wbl}. Similarly, the production of light-baryon resonances has been extensively studied in $\Lambda_c^+ \to pK^-\pi^+$~\cite{Liu:2019dqc,Zhang:2024jby,Cai:2026vnc},  $\Lambda_c^+ \to p\bar{K}^0\pi^0$~\cite{Li:2026lbo}, $\Lambda_c^+ \to \Lambda\eta\pi^+$~\cite{Xie:2016evi,Wang:2022nac,Xie:2017xwx,Lyu:2024qgc,Lyu:2026ack}, $\Lambda_c^+ \to n\bar{K}^0\pi^+$~\cite{Li:2026umb}, $\Lambda_c^+ \to \pi^0\phi p$~\cite{Xie:2017mbe,Xie:2020jlz}, $\Lambda_c^+ \to \Lambda K^0 \pi^+$~\cite{Zhang:2026igc}, $\Lambda_c^+ \to \bar{K}^0\eta p$~\cite{Xie:2017erh,Xie:2020jlz,Li:2024rqb}, and $\Lambda^+_c \to \Lambda K^+ \bar{K}^0$~\cite{Liu:2023jwo}.

In the light baryon sector, an intriguing observation is the two-pole structure of $\Lambda(1405)$~\cite{Oset:1997it,Oller:2000fj,Jido:2003cb,Lu:2022hwm,Xie:2023cej,He:2026mkf}. 
The  hyperon $\Lambda(1405)$ with $J^P = 1/2^-$ is dynamically generated via $\pi \Sigma$ and $\bar{K}N$ coupled-channel interactions in the chiral unitary approach, and has a two-pole structure, consisting of a lower pole above the $\pi\Sigma$ threshold and a dominant $\bar{K}N$ molecular pole below the $\bar{K}N$ threshold~\cite{Oset:1997it,Oller:2000fj,Jido:2003cb,Hyodo:2007jq,Zhuang:2024udv}.
For the low-lying $\Lambda(1670)$ ($J^P=1/2^-$) resonance, its nature remains uncertain and controversial, and no consensus has been reached so far. To interpret its intrinsic properties, several theoretical pictures have been developed in recent decades. The conventional constituent quark model treats $\Lambda(1670)$ as a compact bare three-quark state~\cite{CrystalBall:2001uhc,Zhong:2008km}. Alternatively, the dynamical generation scenario advocates that $\Lambda(1670)$ emerges purely from coupled-channel interactions among the channels of $\bar{K}N$, $\eta\Lambda$, and $\pi\Sigma$~\cite{Miyahara:2015cja,Xie:2016evi,Wang:2022xqc}.
The third viewpoint is a hybrid description in which $\Lambda(1670)$ arises from the superposition of a bare three‑quark component and dynamically generated molecular components~\cite{Liu:2023xvy}.

Several recent studies have investigated the properties of the $\Lambda(1670)$ resonance and revealed that the line shapes of $\Lambda(1670)$ differ across distinct decay channels. A prominent peak corresponding to the $\Lambda(1670)$ resonance is observed at the $\eta\Lambda$ threshold in the decay $\Lambda_c^+ \to \eta\Lambda\pi^+$ by the Belle and BESIII Collaborations~\cite{Belle:2020xku,BESIII:2024mbf}, and subsequent theoretical studies have provided an interpretation for the peak structure of $\Lambda(1670)$ appearing near the threshold in the $\eta\Lambda$ invariant mass distribution~\cite{Xie:2016evi,Lyu:2024qgc,Wang:2022nac,Duan:2024czu}. Meanwhile, the LHCb and Belle Collaborations also reported a cusp at the $\eta\Lambda$ threshold in the $p K^-$ invariant mass spectrum linked to the $\Lambda(1670)$ resonance in the $\Lambda_c^+ \to p K^-\pi^+$ decay~\cite{Belle:2022cbs,LHCb:2022sck}. Further theoretical investigations have elucidated this cusp structure~\cite{Zhang:2024jby,Duan:2024okk}. Most recently, the analysis presented in Ref.~\cite{Li:2026umb} on the decay $\Lambda_c^+ \to n\bar{K}^0\pi^+$ reveals a distinct dip structure in the $n\bar{K}^0$ invariant mass spectrum.

More recently, the BESIII Collaboration has performed a partial wave analysis and measured the invariant mass distributions of the process $\Lambda_c^+ \to p K^+ K^-$~\cite{BESIII:2026dkj}, and the branching fraction of $\mathcal{B}(\Lambda_c^+\to pK^+K^-)$ is updated to $\left(9.94\pm0.65_{\rm stat}\pm0.50_{\rm syst}\right)\times10^{-4}$~\cite{BESIII:2026dkj}. The branching fractions of $\Lambda_c^+ \to p\phi(1020)$, $p f_0(980)$, $\Lambda(1405)K^+$ and $\Lambda(1670)K^+$ are measured and $\Lambda_c^+ \to \Lambda(1405)K^+$ and $\Lambda_c^+ \to \Lambda(1670)K^+$ are observed for the first time in this measurement. The experimental study of the $\Lambda_c^+\to pK^+K^-$ decay was initiated by the CLEO Collaboration in 1996~\cite{CLEO:1995moc}. In 2002, the Belle Collaboration improved the measurements of both $\Lambda_c^+\to pK^+K^-$ and $\Lambda_c^+\to p\phi$ and separately determined the non-$\phi$-contribution to the three-body final state~\cite{Belle:2001hyr}. These results demonstrated that the observed $pK^+K^-$ signal cannot be exhausted by the quasi-two-body decay $\Lambda_c^+\to p\phi$ alone, indicating the presence of additional hadronic dynamics outside the narrow $\phi(1020)$ region.
In 2016, a substantial improvement was achieved by the BESIII Collaboration, which obtained the integrated branching fractions and reported the $K^+K^-$ invariant mass distribution~\cite{BESIII:2016ozn}, which exhibits the prominent narrow $\phi(1020)$ peak together with a noticeable event accumulation near the $K\bar K$ threshold. In 2018, the LHCb Collaboration performed a high-precision measurement of the branching fraction of $\Lambda_c^+\to pK^-K^+$ relative to $\Lambda_c^+\to pK^-\pi^+$, from which the absolute branching fraction of $\Lambda_c^+\to pK^-K^+$ was  determined~\cite{LHCb:2017xtf}. 

Inspired by the invariant mass spectra obtained by the BESIII Collaboration~\cite{BESIII:2016ozn}, the authors of Ref.~\cite{Wang:2020pem} investigated the roles of $a_0(980)$ and $f_0(980)$ in the decays $\Lambda_c^+\to pK^+K^-$ and $\Lambda_c^+\to p\pi^+\pi^-$ within the chiral unitary approach. Nevertheless, their analysis concentrated mainly on the meson--meson subsystem and did not provide a complete description of the $pK^-$ dynamics or of all interference effects over the full three-body phase space.

In the present work, we investigate the singly Cabibbo-suppressed process $\Lambda_c^+ \to p K^+ K^-$ by implementing the $S$-wave pseudoscalar meson-octet baryon interaction within the chiral unitary approach, which dynamically generates the $\Lambda(1405)$ and $\Lambda(1670)$ states. The dynamically generated states $f_0(980)$ and $a_0(980)$ by the $S$-wave pseudoscalar meson-pseudoscalar meson interaction within the chiral unitary approach are also considered. We also include the intermediate contribution from $\phi(1020)$. The primary goal of our analysis is to test whether the proposed mechanism can explain the data and provide theoretical motivation for more precise experimental investigations of this channel.

This article is organized as follows. In Sec.~\ref{Sec:Formalism}, we present the theoretical framework for the $\Lambda_c^+ \to p K^+ K^-$ decay. Numerical results and discussions are presented in Sec.~\ref{Sec:Results}. Finally, a short summary and an outlook are provided in Sec.~\ref{Sec:summary}.

\section{Theoretical Formalism} \label{Sec:Formalism}

\begin{figure*}[htbp]
\centering
\begin{minipage}{0.48\textwidth}
\centering
\includegraphics[scale=0.75]{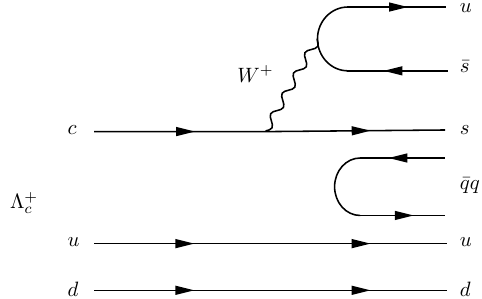}\\
(a)
\end{minipage}
\begin{minipage}{0.48\textwidth}
\centering
\includegraphics[scale=0.75]{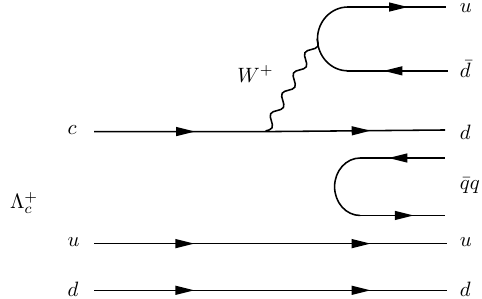}\\
(b)
\end{minipage}

\vspace{1em} 

\begin{minipage}{0.48\textwidth}
\centering
\includegraphics[scale=0.85]{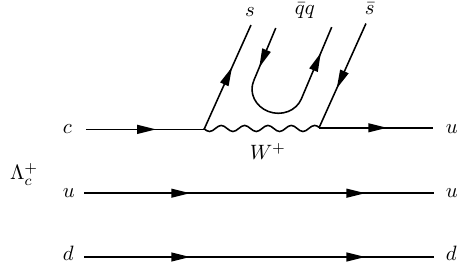}\\
(c)
\end{minipage}
\begin{minipage}{0.48\textwidth}
\centering
\includegraphics[scale=0.85]{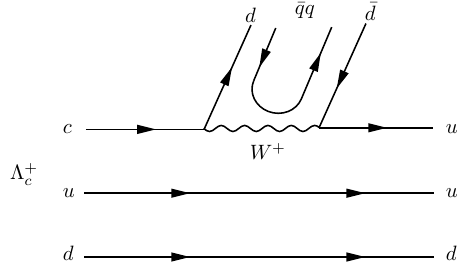}\\
(d)
\end{minipage}

\caption{Quark level diagrams for the process (a) $\Lambda_c^+ \to u\bar{s}s(\bar{q}q)ud$, (b) $\Lambda_c^+ \to u\bar{d}d(\bar{q}q)ud$ via the $W^+$ external emission, (c) $\Lambda_c^+ \to s(\bar{q}q)\bar{s}uud$, and (d) $\Lambda_c^+ \to d(\bar{q}q)\bar{d}uud$ via the $W^+$ internal emission.}
\label{fig:quark_level}
\end{figure*}

For the singly Cabibbo-suppressed process $\Lambda_c^+ \to p K^+ K^-$, we consider the internal and external emission mechanisms at the quark level, as depicted in Fig.~\ref{fig:quark_level}. 
For the external $W^+$ emission mechanism as shown in Fig.~\ref{fig:quark_level}(a), the $c$ quark of the initial $\Lambda_c^+$ weakly decays into an $s$ quark with the emission of a $W^+$ boson, followed by the $W^+$ boson decay into the quark pair $\bar{s}u$. The $sud$ and a $\bar{q}q$ pair created from vacuum hadronize into meson-baryon pairs. Adopting the $\Lambda_c^+$ wave function proposed in Refs.~\cite{Wang:2022nac,Lyu:2024qgc,Li:2026umb,Lyu:2026ack}, we write  the subsequent hadronization process as follows,
\begin{equation}
\begin{aligned}
\Lambda_c^+ &= \frac{1}{\sqrt{2}} c (ud-du) \chi_{MA} \\
&\Rightarrow V_{cs}V_{us} \frac{1}{\sqrt{2}} u\bar{s} \, s(\bar{q}q) (ud-du) \chi_{MA} \\
&\Rightarrow V_{cs}V_{us} K^+ \frac{1}{\sqrt{2}} \sum_i P_{3i} q_i (ud-du) \chi_{MA} \\
&\Rightarrow V_{cs}V_{us} K^+ \frac{1}{\sqrt{2}} \sum_i P_{3i} B_{i3} \chi_{MA} \\
&\Rightarrow V_{cs}V_{us} K^+ \left( \frac{1}{\sqrt{2}} K^- p + \frac{1}{\sqrt{2}} \bar{K}^0 n + \frac{1}{3} \eta \Lambda \right),
\label{eq:ex_a}
\end{aligned}
\end{equation}
where $\chi_{MA}$ represents the mixed antisymmetric spin wave function of the $(ud-du)$ diquark. The pseudoscalar meson matrix $P$ is adopted with the approximate $\eta$-$\eta^{\prime}$ mixing scheme as used in Ref.~\cite{Bramon:1992kr},
\begin{equation}
\begin{aligned}
	P  &= \begin{pmatrix}
        u\bar{u} & u\bar{d} & u\bar{s} \\
        d\bar{u} & d\bar{d} & d\bar{s} \\
        s\bar{u} & s\bar{d} & s\bar{s}
        \end{pmatrix} \\[6pt]
       &= \begin{pmatrix} 
		\dfrac{\eta}{\sqrt{3}}+ \dfrac{\pi^0}{\sqrt{2}}+ \dfrac{\eta^{\prime}}{\sqrt{6}} & \pi^+ & K^+  \\
		\pi^-  &   \dfrac{\eta}{\sqrt{3}}- \dfrac{\pi^0}{\sqrt{2}}+ \dfrac{\eta^{\prime}}{\sqrt{6}}  &  K^0 \\
		K^-  &  \bar{K}^{0}   &    -\dfrac{\eta}{\sqrt{3}}+ \dfrac{\sqrt{6}\eta^{\prime}}{3}
	\end{pmatrix},
\end{aligned}
\end{equation}
where all channels containing the $\eta^{\prime}$ component are neglected due to the large mass of the $\eta^{\prime}$ meson~\cite{Bramon:1992kr,Lyu:2023ppb,Li:2025gvo}. The flavor matrix for the baryon octet is given by
\begin{equation}
\begin{aligned}
B &=\frac{1}{\sqrt{2}}
\begin{pmatrix}
u(ds - sd) & u(su - us) & u(ud - du) \\[3pt]
d(ds - sd) & d(su - us) & d(ud - du) \\[3pt]
s(ds - sd) & s(su - us) & s(ud - du)
\end{pmatrix} \\[6pt]
&=
\begin{pmatrix}
\dfrac{\Sigma^0}{\sqrt{2}}+\dfrac{\Lambda}{\sqrt{6}} & \Sigma^+ & p \\
\Sigma^- & -\dfrac{\Sigma^0}{\sqrt{2}}+\dfrac{\Lambda}{\sqrt{6}} & n \\
\Xi^- & \Xi^0 & -\dfrac{2\Lambda}{\sqrt{6}}
\end{pmatrix}.
\end{aligned}
\end{equation}

Similarly, the weak-decay and hadronization process for the external $W^+$  emission mechanism shown in Fig.~\ref{fig:quark_level}(b) can be expressed as
\begin{equation}
\begin{aligned}
\Lambda_c^+ &= \frac{1}{\sqrt{2}} c (ud-du) \chi_{MA} \\
&\Rightarrow V_{cd} V_{ud} \frac{1}{\sqrt{2}} u\bar{d} \, d(\bar{q}q) (ud-du) \chi_{MA} \\
&\Rightarrow V_{cd} V_{ud}~ \pi^+ \frac{1}{\sqrt{2}} \sum_i P_{2i} q_i (ud-du) \chi_{MA} \\
&\Rightarrow V_{cd} V_{ud}~\pi^+ \left( \frac{1}{\sqrt{2}} \pi^- p + \frac{1}{\sqrt{6}} \eta n - \frac{1}{2} \pi^0 n - \frac{1}{\sqrt{3}} K^0 \Lambda \right).
\label{eq:ex_b}
\end{aligned}
\end{equation}

Next, we investigate the $W^+$ internal emission mechanism presented in Figs.~\ref{fig:quark_level}(c) and \ref{fig:quark_level}(d). For the Feynman diagram shown in Fig.~\ref{fig:quark_level}(c), the $c$ quark of the $\Lambda_c^+$ baryon decays into an $s$ quark and a $W^+$ boson, where the $W^+$ boson further decays into the quark pair $\bar{s}u$. The $s\bar{s}$ and the $\bar{q}q$ created from vacuum fluctuations hadronize into a pair of mesons. Using the wave function of the $\Lambda_c^+$ baryon, the weak-decay and hadronization processes for this mechanism are expressed as follows,
\begin{equation}
\begin{aligned}
\Lambda_c^+ &= \frac{1}{\sqrt{2}} c (ud-du)\chi_{MA} \\
&\Rightarrow V_{cs}V_{us} \frac{1}{\sqrt{2}} s(\bar{q}q)\bar{s}u(ud-du)\chi_{MA} \\
&\Rightarrow V_{cs}V_{us} \sum_i P_{3i}P_{i3} \frac{1}{\sqrt{2}}u(ud-du) \chi_{MA} \\
&\Rightarrow V_{cs}V_{us} \left(\frac{1}{\sqrt{2}} K^+K^- + \frac{1}{\sqrt{2}} K^0\bar{K}^0 + \frac{1}{3\sqrt{2}}\eta\eta  \right) p.
\label{eq:in_c}
\end{aligned}
\end{equation}

For the $W^+$ internal emission mechanism of  Fig.~\ref{fig:quark_level}(d), the generated $d\bar{d}$ quark pair combines with the $\bar{q}q$ pair originating from vacuum fluctuations and hadronizes into a pair of mesons. The weak-decay and hadronization process for this process is explicitly expressed as follows,
\begin{equation}
\begin{aligned}
\Lambda_c^+ &= \frac{1}{\sqrt{2}} c (ud-du) \chi_{MA} \\
&\Rightarrow V_{cd} V_{ud}\frac{1}{\sqrt{2}} d(\bar{q}q)\bar{d}u(ud-du)\chi_{MA} \\
&\Rightarrow V_{cd} V_{ud}\sum_i P_{2i}P_{i2} \frac{1}{\sqrt{2}}u(ud-du) \chi_{MA} \\
&\Rightarrow V_{cd} V_{ud} \left( \frac{1}{\sqrt{2}}\pi^-\pi^+ + \frac{1}{3\sqrt{2}}\eta\eta - \frac{1}{\sqrt{3}}\pi^0\eta \right. \\
&\left. \qquad+ \frac{1}{2\sqrt{2}}\pi^0\pi^0 + \frac{1}{\sqrt{2}}K^0\bar{K}^0 \right) p,
\label{eq:in_d}
\end{aligned}
\end{equation}
where $V_{cs}$, $V_{us}$, $V_{cd}$, and $V_{ud}$ in Eqs.~(\ref{eq:ex_a}), (\ref{eq:ex_b})--(\ref{eq:in_d}) denote the CKM matrix elements for the quark transitions. From the CKM matrix, one has $V_{cs}\simeq V_{ud}$ and $V_{us}\simeq V_{cd}$, so that $V_{cs}V_{us}\simeq V_{ud}V_{cd}$.

\subsection{Production of $\Lambda(1405)$ and $\Lambda(1670)$}

\begin{figure}[htbp]
	\centering
	\begin{minipage}{0.8\columnwidth}
		\centering
		\includegraphics[scale=1]{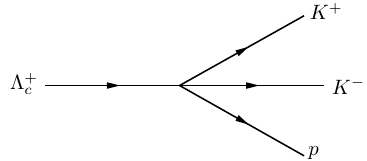}\\
		(a)
	\end{minipage}
	
	\vspace{0.5cm} 
	\begin{minipage}{0.8\columnwidth}
		\centering
		\includegraphics[scale=0.9]{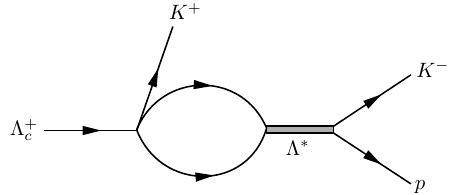}\\
		(b)
	\end{minipage}
	\caption{Hadron-level diagrams for the process $\Lambda_c^+ \to p K^+ K^-$, including (a) tree level diagram and (b) meson-baryon final-state interactions. }
\label{fig:MB}
\end{figure}

In this section, we present the tree-level amplitude for the direct production of the final states $p K^+ K^-$, and the amplitudes for the dynamically generated resonances $\Lambda^*$ [for $\Lambda(1405)$ and $\Lambda(1670)$], respectively depicted in 
Figs.~\ref{fig:MB}(a) and \ref{fig:MB}(b), which are expressed as follows,
\begin{equation}
\begin{aligned}
\mathcal{T}^{\rm tree} = V_{MB}\left( C h_{K^-p} + h_{K^+K^-}    \right),
\label{eq:t_tree}
\end{aligned}
\end{equation}
\begin{equation}
\begin{aligned}
\mathcal{T}^{MB} 
&= V_{MB} \Big\{C\Big[ h_{K^-p}G_{K^-p}(M_{K^-p})t_{K^-p \to K^-p}(M_{K^-p}) \\
&\quad + h_{\bar{K}^0n}G_{\bar{K}^0n}(M_{K^-p})t_{\bar{K}^0n \to K^-p}(M_{K^-p}) \\
&\quad + h_{\eta \Lambda}G_{\eta \Lambda}(M_{K^-p})t_{\eta \Lambda \to K^-p}(M_{K^-p}) \Big] \\
&\quad + h_{K^+K^-}G_{K^-p}(M_{K^-p})t_{K^-p \to K^-p}(M_{K^-p}) \Big\},
\label{eq:t_Lam}
\end{aligned}
\end{equation}
where the parameter $C$ characterizes the relative weight of the internal and external mechanisms at the quark level, and we take $C=3$.\footnote{The value of $C$ is taken to be 3, because the quarks from the $W$ decay in the external emission diagram (for example, the $u$ and $\bar{s}$ of Fig.~\ref{fig:quark_level}(a) have three choices of the colors (we take $N_c = 3$), while the quarks from the $W$ decay in the internal emission diagram (for example, the $u$ and $\bar{s}$ of Fig.~\ref{fig:quark_level}(c) have the fixed colors. }
 $V_{MB}$ is the production strength of $\Lambda(1405)$ and $\Lambda(1670)$, which also absorbs the CKM matrix elements $V_{cs}V_{us}$ and $V_{ud}V_{cd}$. $h_i$ originate from the flavor structure of Eqs.~(\ref{eq:ex_a}) and (\ref{eq:in_c}), which are taken as follows
\begin{equation}
\begin{aligned}
h_{K^-p}=\frac{1}{\sqrt{2}} , h_{\bar{K}^0n}=\frac{1}{\sqrt{2}} , h_{\eta \Lambda}=\frac{1}{3} , h_{K^+K^-}=\frac{1}{\sqrt{2}} .
\label{eq:hi_MB}
\end{aligned}
\end{equation}

The loop function ${G}_i$ of the meson-baryon system in Eq.~(\ref{eq:t_Lam}) is given by the dimensional regularization method,
  \begin{eqnarray}
\begin{aligned}
G_{l}  =&i\int\frac{d^4q}{(2\pi)^4}\frac{2M_i}{(P-q)^2-M_i^2+i\epsilon}\frac{1}{q^2-m_i^2+i\epsilon} \\
		 =&\frac{2M_l}{16 \pi^2}\left\{a_l(\mu)+\ln \frac{M_l^2}{\mu^2}+\frac{s+m_l^2-M_l^2}{2 s} \ln \frac{m_l^2}{M_l^2}\right. \\
		& +\frac{|\vec{q}\,|}{\sqrt{s}}\left[\ln \left(s-\left(M_l^2-m_l^2\right)+2 |\vec{q}\,| \sqrt{s}\right)\right. \\
		& +\ln \left(s+\left(M_l^2-m_l^2\right)+2 |\vec{q}\,| \sqrt{s}\right) \\
		& -\ln \left(-s+\left(M_l^2-m_l^2\right)+2 |\vec{q}\,| \sqrt{s}\right) \\
		& \left.\left.-\ln \left(-s-\left(M_l^2-m_l^2\right)+2 |\vec{q}\,| \sqrt{s}\right)\right]\right\},
\end{aligned}
 \label{eq:Gi},
 \end{eqnarray}
where $M_{i}$ and $m_{i}$ are the masses of baryon and meson of $i$-th coupled channel, respectively. $P$ is the four-momentum of the meson-baryon system, and $q$ is the four-momentum of meson in the center-of-mass frame. 

 The coupled-channels scattering matrices $t_{MB \to K^- p}$ are obtained by solving the Bethe-Salpeter equation,
\begin{eqnarray}
\begin{aligned}
T=[1-VG]^{-1}V,
 \label{eq:BS}
\end{aligned}
 \end{eqnarray}
and the transition potential $V_{ij}$ is taken from Ref.~\cite{Oset:1997it},
\begin{eqnarray}
\begin{aligned}
V_{ij}& =-\frac{C_{ij}}{4f^2}\left(2\sqrt{s}-M_i-M_j\right)  \\
&\times\left(\frac{M_i+E_i}{2M_i}\right)^{1/2}\left(\frac{M_j+E_j}{2M_j}\right)^{1/2},
 \label{eq:Vij}
\end{aligned}
 \end{eqnarray}
where $E_i$ and $M_i$ are the energy and mass of the baryon in the $i$-th channel, and the coefficients $C_{ij}$ that reflect the SU(3) flavor symmetry are given in Table~I of Ref.~\cite{Oset:1997it}. The coupling $f$ is the pseudoscalar decay constant, for which we use $f=1.15f_{\pi},~f_{\pi}=93\mathrm{~MeV}$.
The loop function $G$ in Eq.~(\ref{eq:BS}) is evaluated using the dimensional regularization method, as shown in Eq.~(\ref{eq:Gi}). 
We adopt the following subtraction constants $a_i$ from Ref.~\cite{Oset:2001cn}, except the value of $a_{K \Xi}$, which is taken from Ref.~\cite{Zhang:2024jby},
\begin{eqnarray}
\begin{aligned}
a_{\bar{K}N} &= -1.84, & a_{\pi\Sigma} &= -2.00, &a_{\pi\Lambda} &= -1.83, \\
  a_{\eta\Lambda} &= -2.25, & a_{\eta\Sigma} &= -2.38, & a_{K\Xi} &= -2.776.
\end{aligned}
\end{eqnarray}

\subsection{Production of $f_0(980)$ and $a_0(980)$ }

\begin{figure}[htbp]
		\centering
		\includegraphics[scale=1]{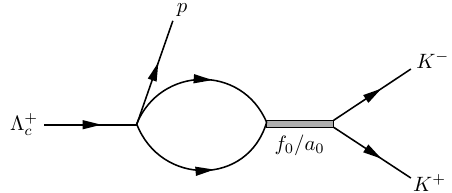}
 \caption{Hadron-level diagram for the process $\Lambda_c^+ \to p K^+ K^-$ with meson–meson final-state interactions, where $f_0/a_0$ denote $f_0(980)$ or $a_0(980)$, respectively. }
\label{fig:MM}
\end{figure}

The process $\Lambda_c^+ \to p K^+K^- $ can proceed through the $S$-wave pseudoscalar meson-pseudoscalar meson final-state interactions illustrated in Fig.~\ref{fig:MM}. Accordingly, the corresponding amplitude for the diagram in Fig.~\ref{fig:MM} reads

\begin{equation}
\begin{aligned}
&\quad \quad \mathcal{T}^{MM} \\
&= V_{MM} \Bigg\{ Ch_{K^-p}G_{K^+K^-}(M_{K^+K^-})t_{K^+K^- \to K^+K^-}(M_{K^+K^-}) \\
&\quad + Ch_{\pi^-p}G_{\pi^+\pi^-}(M_{K^+K^-})t_{\pi^+\pi^- \to K^+K^-}(M_{K^+K^-}) \\
&\quad + \Big[ h_{K^+K^-}G_{K^+K^-}(M_{K^+K^-})t_{K^+K^- \to K^+K^-}(M_{K^+K^-}) \\
&\quad + h_{K^0 \bar{K}^0}G_{K^0 \bar{K}^0}(M_{K^+K^-})t_{K^0 \bar{K}^0 \to K^+K^-}(M_{K^+K^-}) \\
& \quad + 2 \times \frac{1}{2} h_{\eta \eta}G_{\eta \eta}(M_{K^+K^-})\tilde{t}_{\eta \eta \to K^+K^-}(M_{K^+K^-}) \Big] \\
& \quad + \Big[  h_{\pi^+\pi^-}G_{\pi^+\pi^-}(M_{K^+K^-})t_{\pi^+\pi^- \to K^+K^-}(M_{K^+K^-}) \\
&\quad + 2 \times \frac{1}{2} h_{\eta \eta}G_{\eta \eta}(M_{K^+K^-})\tilde{t}_{\eta \eta \to K^+K^-}(M_{K^+K^-}) \\
& \quad + h_{\pi^0 \eta}G_{\pi^0 \eta}(M_{K^+K^-})t_{\pi^0 \eta \to K^+K^-}(M_{K^+K^-}) \\
&\quad + 2 \times \frac{1}{2} h_{\pi^0 \pi^0}G_{\pi^0 \pi^0}(M_{K^+K^-})\tilde{t}_{\pi^0 \pi^0 \to K^+K^-}(M_{K^+K^-}) \\
&\quad + h_{K^0 \bar{K}^0}G_{K^0 \bar{K}^0}(M_{K^+K^-})t_{K^0 \bar{K}^0 \to K^+K^-}(M_{K^+K^-}) \Big]
\Bigg\}.
\end{aligned}
\end{equation}
Based on the isospin of each reaction channel, we perform a  rearrangement of the above expression and obtain
\begin{equation}
\begin{aligned}
&\mathcal{T}^{MM}= V_{MM} \Bigg[ \Big( Ch_{\pi^- p} + h_{\pi^+ \pi^-} \Big) \\ &\quad \times G_{\pi^+ \pi^-}(M_{K^+K^-})t_{\pi^+\pi^- \to K^+K^-}(M_{K^+K^-}) \\
&\quad + h_{K^+K^-}G_{K^+K^-}(M_{K^+K^-})t_{K^+K^- \to K^+K^-}(M_{K^+K^-}) \\
&\quad + h_{K^0 \bar{K}^0}G_{K^0 \bar{K}^0}(M_{K^+K^-})t_{K^0 \bar{K}^0 \to K^+K^-}(M_{K^+K^-}) \\
&\quad + 2 \times 2 \times \frac{1}{2} h_{\eta \eta}G_{\eta \eta}(M_{K^+K^-})\tilde{t}_{\eta \eta \to K^+K^-}(M_{K^+K^-}) \\
& \quad + 2 \times \frac{1}{2} h_{\pi^0 \pi^0}G_{\pi^0 \pi^0}(M_{K^+K^-})\tilde{t}_{\pi^0 \pi^0 \to K^+K^-}(M_{K^+K^-}) \Bigg] \\
&\quad +V_{MM} \Bigg[ Ch_{K^-p}G_{K^+K^-}(M_{K^+K^-})t_{K^+K^- \to K^+K^-}(M_{K^+K^-}) \\
&\quad + h_{\pi^0 \eta}G_{\pi^0 \eta}(M_{K^+K^-})t_{\pi^0 \eta \to K^+K^-}(M_{K^+K^-}) \\
&\quad + h_{K^0 \bar{K}^0}G_{K^0 \bar{K}^0}(M_{K^+K^-})t_{K^0 \bar{K}^0 \to K^+K^-}(M_{K^+K^-}) \Bigg],
\label{eq:t_MM}
\end{aligned}
\end{equation}
where the first term only carries the amplitude arising from isospin $I=0$, and the second term includes both $I=0$ and $I=1$ contributions generated by the dynamical mechanisms of Figs.~\ref{fig:quark_level}(a) and  \ref{fig:quark_level}(d). The coefficients $h_i$ correspond to the production weights for individual components in the primary decay, which are obtained from Eqs.~(\ref{eq:ex_a}), (\ref{eq:ex_b})-(\ref{eq:in_d}),
\begin{equation}
\begin{aligned}
&h_{K^-p}= h_{\pi^- p} = h_{K^0 \bar{K}^0} = h_{K^+K^-} = h_{\pi^+ \pi^-} = \frac{1}{\sqrt{2}}, \\
&h_{\eta \eta}=\frac{1}{3\sqrt{2}},\quad h_{\pi^0 \eta}=-\frac{1}{\sqrt{3}} ,\quad h_{\pi^0 \pi^0}=\frac{1}{2\sqrt{2}} .
\label{eq:hi_MM}
\end{aligned}
\end{equation}

In Eq.~(\ref{eq:t_MM}), a factor of 2 is introduced to account for the two distinct ways of matching the two identical particles contained in the operators of Eqs.~(\ref{eq:in_c}) and (\ref{eq:in_d}) to the produced two-meson systems $\pi^0\pi^0$ and $\eta\eta$. Meanwhile, a factor of ${1}/{2}$ arises for intermediate loops involving a pair of identical mesons \cite{Liang:2014ama,Liang:2015qva,Wang:2020pem}.
$V_{MM}$ denotes the coupling strength of the weak production vertex and can be regarded as a constant. By taking $G_{K^0\bar{K}^0} = G_{K^+K^-}$ and adopting the treatment scheme given in Refs.~\cite{Wang:2020pem,Liang:2015qva}, one may rearrange the relevant pieces of the transition amplitudes $t_{K^+K^-\to K^+K^-}$ and $t_{K^0\bar{K}^0\to K^+K^-}$ appearing in the second term of Eq.~(\ref{eq:t_MM}), which leads to
\begin{equation}
\begin{aligned}
&\quad C h_{K^-p} G_{K^+K^-} t_{K^+K^-\to K^+K^-} + h_{K^0\bar{K}^0}G_{K^0\bar{K}^0} t_{K^0\bar{K}^0\to K^+K^-} \\
&=  C h_{K^0\bar{K}^0} G_{K^0\bar{K}^0} t_{K^+K^-\to K^+K^-} + h_{K^0\bar{K}^0}G_{K^0\bar{K}^0}  t_{K^0\bar{K}^0\to K^+K^-} \\
&= h_{K^0\bar{K}^0}  G_{K^0\bar{K}^0} \Big( Ct_{K^+K^-\to K^+K^-} + t_{K^0\bar{K}^0\to K^+K^-}  \Big) \\
&= h_{K^0\bar{K}^0} G_{K^0\bar{K}^0} \Bigg[ \frac{1+C}{2} \big(t_{K^0\bar{K}^0\to K^+K^-} + t_{K^+K^-\to K^+K^-}\big) \\
&\quad + \frac{1-C}{2} \big(t_{K^0\bar{K}^0\to K^+K^-} - t_{K^+K^-\to K^+K^-}\big) \Bigg],
\label{eq:GKK}
\end{aligned}
\end{equation}
where the first two terms are in $I = 0$ while the last two terms are in $I = 1$. From Eqs.~(\ref{eq:ex_a}) and (\ref{eq:in_d}), we obtain $h_{K^0\bar{K}^0} = h_{K^+K^-} = {1}/{\sqrt{2}}$, so $h_{K^0\bar{K}^0}$ can be substituted for $h_{K^+K^-}$ in the present expression. Then, the amplitude from Eq.~(\ref{eq:t_MM}) can be rewritten as
\begin{equation}
\begin{aligned}
&\mathcal{T}^{MM} 
= V_{MM} \Bigg\{ \Big(Ch_{\pi^- p} + h_{\pi^- \pi^+}\Big)G_{\pi^+\pi^-}t_{\pi^+\pi^- \to K^+K^-} \\
&+ \frac{3+C}{2} h_{K^0\bar{K}^0}  G_{K^0\bar{K}^0} \Big(t_{K^0\bar{K}^0\to K^+K^-} + t_{K^+K^-\to K^+K^-}\Big) \\
&  + 2 h_{\eta \eta}G_{\eta \eta}\tilde{t}_{\eta \eta \to K^+K^-}  + h_{\pi^0 \pi^0}G_{\pi^0 \pi^0}\tilde{t}_{\pi^0 \pi^0 \to K^+K^-}
\Bigg\} \\
&  + V_{MM} \Bigg\{ h_{\pi^0 \eta}G_{\pi^0 \eta}t_{\pi^0 \eta \to K^+K^-}  \\
&  + \frac{1-C}{2} h_{K^0\bar{K}^0}  G_{K^0\bar{K}^0} \Big(t_{K^0\bar{K}^0\to K^+K^-} - t_{K^+K^-\to K^+K^-}\Big)
\Bigg\} \\
&=t^{MM}_{I=0} + t^{MM}_{I=1},
\label{eq:re_t_MM}
\end{aligned}
\end{equation}
where the terms $t^{MM}_{I=0}$ and $t^{MM}_{I=1}$ correspond to the contributions from the $I=0$ and $I=1$ channels, respectively. For channels with two identical mesons, we adopt the relations $\tilde{t}_{\eta\eta\to i} = \sqrt{2}\,t_{\eta\eta\to i}$ and $\tilde{t}_{\pi^0\pi^0\to j} = \sqrt{2}\,t_{\pi^0\pi^0\to j}$ \cite{Dias:2016gou}. 
$t_{MM\to K^+K^-}$ denotes the scattering matrix for meson–meson interactions, which can be obtained by solving the Bethe-Salpeter equation presented in Eq.~(\ref{eq:BS}). $V_{i \to j}$ is the matrix of the interaction kernel corresponding to the tree-level transition amplitudes obtained from phenomenological Lagrangians~\cite{Oller:1997ti}. This matrix describes five coupled channels $\pi^+\pi^-$, $\pi^0\pi^0$, $K^+K^-$, $K^0\bar{K}^0$, and $\eta\eta$ for isospin $I=0$, while three channels $K^+K^-$, $K^0\bar{K}^0$, and $\pi^0\eta$ are taken into account for $I=1$. The explicit expressions for $I=0$ and $I=1$ are presented in Eqs.~(19) and (20) of Ref.~\cite{Wang:2020pem}.

For the meson-meson system, the loop function $G$ is given by the cutoff method with the cutoff parameter $q_{\text{max}} = 600$~MeV~\cite{Wang:2020pem,Li:2025gvo,Xie:2014tma}, and it can be expressed as
\begin{equation}
\begin{aligned}
G &= i \int \frac{d^4 q}{(2\pi)^4} \frac{1}{(P - q)^2 - m_1^2 + i\epsilon} \frac{1}{q^2 - m_2^2 + i\epsilon} \\
&= \int_{0}^{q_{\text{max}}} \frac{\mathbf{q}^2 d\mathbf{q}}{(2\pi)^2} \frac{\omega_1 + \omega_2}{\omega_1 \omega_2 } \frac{1}{ s' - (\omega_1 + \omega_2)^2 + i\epsilon},
\end{aligned}
\label{eq:loopG_23}
\end{equation}
where $s'$ is the invariant mass squared of the meson-meson pair, and the meson energies read $\omega_i = \sqrt{\mathbf{q}^2 + m_i^2}$.

To obtain a physical $K^+K^-$ invariant mass spectrum, we compute the $Gt$ combinations from Eq.~(\ref{eq:re_t_MM}) up to the invariant mass $M_{\text{cut}} = 1100$~MeV. For invariant masses exceeding $M_{\text{cut}}$, we suppress $Gt$ via an exponential smoothing factor to ensure a gradual falloff at large $M_{K^+K^-}> M_{\text{cut}}$, following the prescription of Refs.~\cite{Zhang:2024myn,Debastiani:2016ayp},
\begin{equation}
\begin{aligned}
Gt(M_{K^+K^-}) = Gt(M_{\text{cut}})e^{-\alpha\left(M_{K^+K^-}-M_{\text{cut}}\right)}.
\end{aligned}
\label{eq:cut}
\end{equation}
We adopt $\alpha = 0.0054\ \text{MeV}^{-1}$ for this smoothing parameter, as recommended in Ref.~\cite{Debastiani:2016ayp}.

\begin{figure}[htbp]
		\centering
		\includegraphics[scale=1]{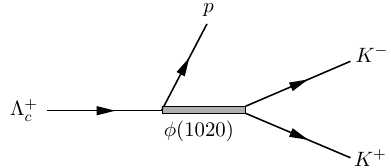}
 \caption{Hadron-level diagram for the process $\Lambda_c^+ \to p \phi(1020) \to p K^+ K^-$. }
\label{fig:phi1020}
\end{figure}

\subsection{Contribution from the $\phi(1020)$}
\label{subsec:phi1020}

We also take into account the contribution from the intermediate vector
meson $\phi(1020)$, as depicted in Fig.~\ref{fig:phi1020}.
Following Ref.~\cite{Klingl:1996by}, the $\phi K^+K^-$ vertex is described by the standard derivative coupling for a vector meson and two pseudoscalar mesons,

\begin{equation}
\mathcal{L}_{\phi KK}
= -g_{\phi KK}\,
  \phi^{\mu}\left(K^+\partial_{\mu}K^- - K^-\partial_{\mu}K^+\right).
\label{eq:app-Lag}
\end{equation}
The angular dependence associated with the different $\phi$ helicities follows the Jacob-Wick helicity formalism~\cite{Jacob:1959at} (See Appendix~\ref{app:phi}).
To extend the purely helicity-zero amplitude while preserving its original momentum dependence, we employ the following minimal
spin-dependent parametrization:
\begin{equation}
\begin{aligned}
\mathcal{T}^{\phi}_{s_p s_{\Lambda_c}}
=
\mathcal{C}_{\phi}(M_{K^+K^-})\,
\chi_{s_p}^{\dagger}
\Big[
\cos\theta\,\mathbf{1}
+ i r_T\sin\theta\,
\bm{\sigma}\cdot\widehat{\bm n}
\Big]
\chi_{s_{\Lambda_c}},
\end{aligned}
\label{eq:phi_full_amplitude}
\end{equation}
where
\begin{equation}
\begin{aligned}
\mathcal{C}_{\phi}(M_{K^+K^-})
= \frac{V_{\phi(1020)}
|\bm p_{K^+}^{\,*}|
|\bm p_p^{\,*}|}{ D_{\phi}(M_{K^+K^-}) },
\end{aligned}
\label{eq:phi_common_amplitude}
\end{equation}
with the Breit-Wigner denominator
\begin{equation}
\begin{aligned}
D_{\phi}(M_{K^+K^-}) =
M_{K^+K^-}^2-m_{\phi}^{2}+i m_{\phi} \widetilde{\Gamma}_{\phi} \left(M_{K^+K^-}\right).
\end{aligned}
\label{eq:phi_bw_denominator}
\end{equation}
Here, $\chi_{s_{\Lambda_c}}$ and $\chi_{s_p}$ denote the Pauli spinors of the initial $\Lambda_c^+$ and final proton, respectively. $\bm{\sigma}$ is the Pauli spin operator. The unit vector normal to the decay plane is defined by
\begin{equation}
\begin{aligned}
\widehat{\bm n}
={}&
\frac{
\bm p_p^{\,*}
\times
\bm p_{K^+}^{\,*}
}{
|\bm p_p^{\,*}|
|\bm p_{K^+}^{\,*}|
\sin\theta
}.
\end{aligned}
\label{eq:phi_normal_vector}
\end{equation}

The first term in Eq.~\eqref{eq:phi_full_amplitude} is spin independent and represents the helicity-zero component, which produces the $\cos\theta$ angular dependence. The second term is spin dependent and represents the transverse component, which produces the $\sin\theta$ dependence. The parameter $r_T$ specifies the transverse amplitude relative to the longitudinal one. 

For an unpolarized $\Lambda_c^+$ and without measuring the proton polarization, the spin-averaged squared amplitude becomes
\begin{equation}
\begin{aligned}
\left|\mathcal{T}^{\phi(1020)}\right|^2
&\equiv
\frac{1}{2}
\sum_{s_{\Lambda_c},s_p}
\left|
\mathcal{T}^{\phi}_{s_p s_{\Lambda_c}}
\right|^2
\\
&=\left|\mathcal{C}_{\phi}(M_{K^+K^-})\right|^2
\left[\cos^2\theta+|r_T|^2\sin^2\theta\right]. 
\end{aligned}
\label{eq:phi_spin_averaged_intensity}
\end{equation}
The interference between the two terms in
Eq.~\eqref{eq:phi_full_amplitude} vanishes after the spin summation because
\begin{equation}
\operatorname{Tr}
\left( \bm{\sigma}\cdot\widehat{\bm n} \right)=0.
\end{equation}


The transverse parameter can also be related to the helicity amplitudes of the weak decay $\Lambda_c^+\to p\phi$ through
\begin{equation}
\begin{aligned}
|r_T|^2=\frac{|H_{+\frac{1}{2},+1}|^2+|H_{-\frac{1}{2},-1}|^2}
{2\Big(|H_{+\frac{1}{2},0}|^2+|H_{-\frac{1}{2},0}|^2\Big)}.
\end{aligned}
\label{eq:phi_rT_helicity}
\end{equation}
Accordingly, $r_T$ is determined by the weak production dynamics and is not fixed by the quantum numbers of the $\phi(1020)$ alone. In the present analysis, $r_T$ can be treated as a free parameter.

$|\bm p_{K^+}^{\,*}|$ and $|\bm p_{p}^{\,*}|$ are the momenta of $K^+$ and $p$ in the $K^+ K^-$ rest frame, and $\theta$ is the angle between $K^+$ and $p$ in the center-of-mass frame of the $K^+ K^-$ system, which are given by
\begin{eqnarray}
|\bm p_{K^+}^{\,*}|=\frac{\lambda^{1/2}(M_{K^+ K^-}^{2},m_{K^{+}}^{2},m_{K^-}^{2})}{2M_{K^+ K^-}},
\end{eqnarray}
\begin{eqnarray}
|\bm p_{p}^{\,*}|=\frac{\lambda^{1/2}(M_{\Lambda_c^+}^{2},M_{p}^{2},M_{K^+K^-}^{2})}{2M_{K^+ K^-}},
\end{eqnarray}
\begin{eqnarray}
\text{cos}\theta=\frac{M_{pK^-}^2-M_{\Lambda_c^+}^2 - m_{K^+}^2 + 2P_{\Lambda_c^+}^0P_{K^+}^0}{2|\bm p_{K^+}^{\,*}||\bm p_{p}^{\,*}|},
\end{eqnarray}
with the K\"all\'en function $\lambda(x,y,z) = x^2 + y^2 + z^2 - 2xy - 2yz - 2xz$.
In the $K^+ K^-$ rest frame, $\bm{p}_{\Lambda_c^+}^{\,*} = \bm{p}_p^{\,*}$, $\bm{p}_{K^+}^{\,*} = -\bm{p}_{K^-}^{\,*}$, and the $\Lambda_c^+$ and $K^+$ energies are
\begin{align*}
P_{\Lambda_c^+}^0 &= \sqrt{M_{\Lambda_c^+}^2 + |\bm{p}_{\Lambda_c^+}^{\,*}|^2} = \sqrt{M_{\Lambda_c^+}^2 + |\bm{p}_p^{\,*}|^2}, \\
P_{K^+}^0 &= \sqrt{m_{K^+}^2 + |\bm{p}_{K^+}^{\,*}|^2}.
\end{align*}

Since $\phi(1020)$ decays to $K^+K^-$  via a $P$-wave, the running width has the characteristic cubic momentum dependence. Following Ref.~\cite{Wang:2020pem}, we use
\begin{equation}
\begin{aligned}
\widetilde{\Gamma}_{\phi}
\left(M_{K^+K^-}\right)
={}&
\Gamma_{\phi}\left[\frac{\sqrt{M_{K^+K^-}^{2}-4m_K^2}}{\sqrt{m_{\phi}^{2}-4m_K^2}}\right]^3 .
\end{aligned}
\label{eq:phi_running_width}
\end{equation}

The effective coupling $V_{\phi(1020)}$, which is determined by fitting to experimental data, absorbs the weak production coupling, the $\phi K^+K^-$ coupling, and convention-dependent normalization factors. The parameters $m_{\phi}$ and $\Gamma_{\phi}$ denote the mass and width of the $\phi(1020)$, respectively. 
In the present work, we adopt two fitting setups: in Model A, the mass and width of the $\phi(1020)$ are fixed at $m_{\phi}=1019.46\ \mathrm{MeV}$ and $\Gamma_{\phi}=4.25\ \mathrm{MeV}$~\cite{ParticleDataGroup:2024cfk}; in Model B, $m_{\phi}$ and $\Gamma_{\phi}$ are treated as free parameters and extracted from fits to the experimental data.

\subsection{Invariant mass distributions} 

Using the formalism described above, we write the modulus squared of the total decay amplitude for the $\Lambda_c^+ \to p K^+ K^-$ process as
\begin{equation}
\begin{aligned}
\Big|\mathcal{T}^\text{Total}\big|^2 = \Big| \mathcal{T}^\text{tree} + \mathcal{T}^\text{MB} e^{i\phi_1\pi} +  \mathcal{T}^\text{MM} e^{i\phi_2\pi}\Big|^2  + \Big| \mathcal{T}^{\phi(1020)} \Big|^2,
\end{aligned}
\end{equation}
where $\phi_1$ and $\phi_2$ are the phases of the meson-baryon interaction and meson-meson interaction, respectively.
The double differential decay width for the process $\Lambda_c^+ \to pK^+K^-$  reads
\begin{equation}
\begin{aligned}
\frac{d^{2}\Gamma}{dM_{K^-p}dM_{K^+K^-}}=\frac{1}{(2\pi)^{3}}\frac{M_{p}}{2M_{\Lambda_{c}^{+}}^{2}}|\mathcal{T}^\text{Total}|^{2}M_{K^-p}M_{K^+K^-}.
\end{aligned}
\end{equation}

For a given value of $M_{12}$, the range of $M_{23}$ is determined according to the Review of Particle Physics (RPP)~\cite{ParticleDataGroup:2024cfk},
\begin{align}
	&\left(M_{23}^2\right)_{\min}=\left(E_2^*+E_3^*\right)^2-\left(\sqrt{E_2^{* 2}-m_2^2}+\sqrt{E_3^{* 2}-m_3^2}\right)^2, \nonumber\\
	&\left(M_{23}^2\right)_{\max}=\left(E_2^*+E_3^*\right)^2-\left(\sqrt{E_2^{* 2}-m_2^2}-\sqrt{E_3^{* 2}-m_3^2}\right)^2, \label{eq:limit}
\end{align}
where $E_{2}^*$ and $E_{3}^*$ are the energies of particles 2 and 3 in the $M_{12}$ rest frame, which are written as
\begin{eqnarray}
E_{2}^{*}=\frac{M_{12}^2-m_1^2+m_2^2}{2M_{12}},\\E_{3}^{*}=\frac{M_{\Lambda_{c}^{+}}^{2}-M_{12}^{2}-m_{3}^{2}}{2M_{12}},
\label{eq:e23}
\end{eqnarray}
while $m_1$, $m_2$, and $m_3$ are the masses of particles 1, 2, and 3, respectively. Permutation of the indices allows us to evaluate all three mass distributions, using $M_{12}$ and $M_{23}$ as independent variables, and the property $M_{12}^2+M_{13}^2+M_{23}^2=M_{\Lambda_c^+}^2+M_p^2+m_{K^+}^2+m_{K^-}^2$ to get $M_{13}$ from them. The masses and widths of the particles are taken from the RPP~\cite{ParticleDataGroup:2024cfk}. 

To compare the theoretical invariant mass distributions with experimental data, the finite detector resolution should be taken into account. This effect is particularly relevant for the narrow $\phi(1020)$ resonance, whose natural width is comparable to the experimental $K^+K^-$ mass resolution, and the bin width of the measured $K^+K^-$ invariant mass distribution is larger than the natural width of $\phi(1020)$. The detector smearing causes events generated at a given true invariant mass to migrate into neighboring reconstructed mass bins, thereby modifying the position, height, and width of the observed $\phi(1020)$ peak. 

We first obtain the truth-level $K^+K^-$ invariant mass distribution by integrating the double differential decay width over the kinematically allowed range of $M_{K^-p}$,
\begin{equation}
\begin{aligned}
F(M_{K^+K^-})&=\frac{d\Gamma}{dM_{K^+K^-}}\\
&=
\int_{M_{K^-p}^{\min}}^{M_{K^-p}^{\max}}
\frac{d^2\Gamma}{dM_{K^-p}\,dM_{K^+K^-}}\,
dM_{K^-p},
\end{aligned}
\end{equation}
The experimental mass resolution function is parametrized as the weighted sum of two Gaussian functions,
\begin{equation}
R(M_{\rm rec},M)
=
\sum_{i=1}^{2}
\frac{\alpha_i}{\sqrt{2\pi}\sigma_i}
\exp\left[
-\frac{\left(M_{\rm rec}-M-\delta_i\right)^2}
{2\sigma_i^2}
\right],
\end{equation}
where $M$ and $M_{\rm rec}$ denote the true and reconstructed $K^+K^-$ invariant masses, respectively. The relative Gaussian weights satisfy~\cite{BESIII:2026dkj}
\begin{equation}
\alpha_1+\alpha_2=1,
\qquad
\alpha_1=\frac{1}{1+8.38},
\qquad
\alpha_2=\frac{8.38}{1+8.38}.
\end{equation}
The corresponding mean offsets and standard deviations are
\begin{equation}
\begin{aligned}
&\delta_1=11.5~\mathrm{MeV},
\qquad
\sigma_1=3.80~\mathrm{MeV},\\
&\delta_2=0.60~\mathrm{MeV},
\qquad
\sigma_2=1.69~\mathrm{MeV}.
\end{aligned}
\end{equation}
The detector-smeared distribution is then obtained by folding the truth-level spectrum with the resolution function,
\begin{equation}
F_{\rm conv}(M_{\rm rec})=\int_{M_{K^+K^-}-3{\sigma}_1}^{M_{K^+K^-}+3{\sigma}_1}F(M)\,R(M_{\rm rec},M)\,dM.
\end{equation}

\begin{table}[htbp]
  \centering
  \caption{Values of the fitted parameters. The units of $m_\phi$ and $\Gamma_\phi$ are MeV, and the other parameters are dimensionless.}
  \renewcommand{\arraystretch}{1.8}
  \setlength{\tabcolsep}{8pt}
  \begin{tabular}{l c c}
    \toprule
    \midrule
    Parameter & Model A & Model B \\
    \midrule
    $V_{MB}$ & $(12.48\pm1.25)\times 10^{-2}$ & $(12.55 \pm 1.28)\times 10^{-2}$ \\
    $V_{MM}$ & $(8.08\pm2.35)\times 10^{-2}$ & $(7.96 \pm 2.56)\times 10^{-2}$ \\
    $V_{\phi(1020)}$ & $(5.68\pm0.50)\times 10^{-2}$ & $(5.96 \pm 0.91)\times 10^{-2}$ \\
    $r_T$ & $1.02\pm0.12$ & $1.24 \pm 0.21$ \\
    $\phi_1$ & $0.32\pm0.05$ & $0.32 \pm 0.05$ \\
    $\phi_2$ & $-0.59\pm0.12$ & $3.41 \pm 0.12$ \\
    $m_{\phi(1020)}$ & $-$ & $1018.47 \pm 0.37$ \\
    $\Gamma_{\phi(1020)}$ & $-$ & $5.98 \pm 1.20$ \\
    $\chi^2/\mathrm{d.o.f.}$ & $1.28$ & $1.06$ \\
    \midrule
    \bottomrule
  \end{tabular}
  \label{tab:fit_params}
  \renewcommand{\arraystretch}{1.0}
\end{table}

\section{Numerical results and discussion} \label{Sec:Results}

\begin{figure}[htbp]
\centering
\begin{minipage}{1\columnwidth}
\centering
\begin{overpic}[scale=0.65]{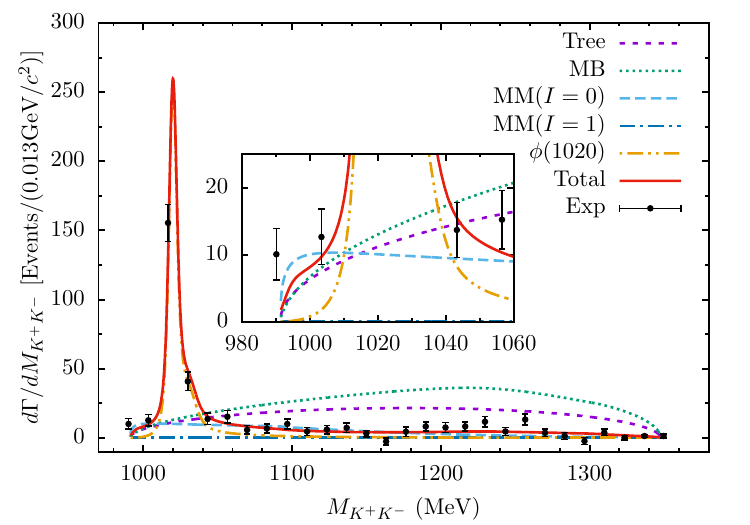}
    \put(16,60){\large (a)}
\end{overpic}
\end{minipage}

\vspace{0.5cm}
\begin{minipage}{1\columnwidth}
\centering
\begin{overpic}[scale=0.65]{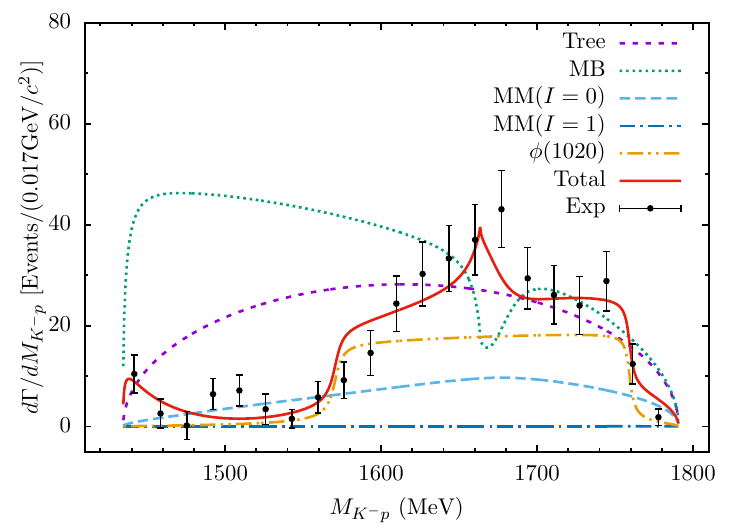}
    \put(15,60){\large (b)}
\end{overpic}
\end{minipage}

\vspace{0.5cm}
\begin{minipage}{1\columnwidth}
\centering
\begin{overpic}[scale=0.65]{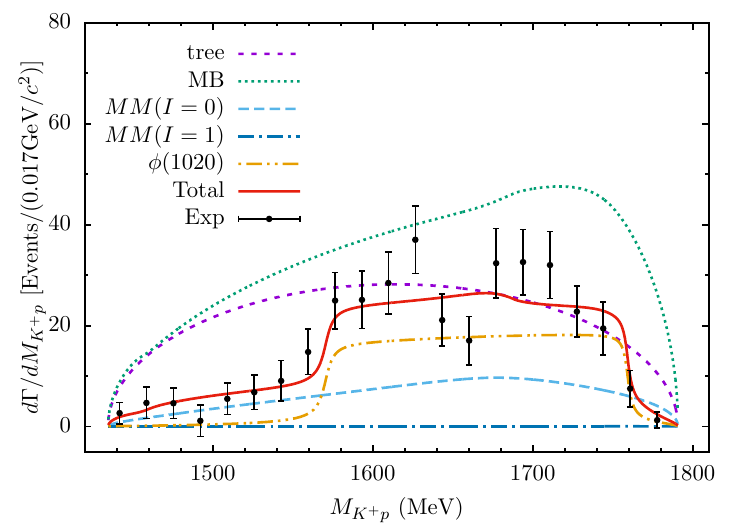}
    \put(85,60){\large (c)}
\end{overpic}
\end{minipage}
\caption{$K^+K^-$ (a), $pK^-$ (b), and $pK^+$ (c) invariant mass distributions of the $\Lambda_c^+ \to pK^+K^-$ decay with the fitted parameters of Model A.}
\label{fig:with_para6}
\end{figure}

\begin{figure}[htbp]
\centering
\begin{minipage}{1\columnwidth}
\centering
\begin{overpic}[scale=0.65]{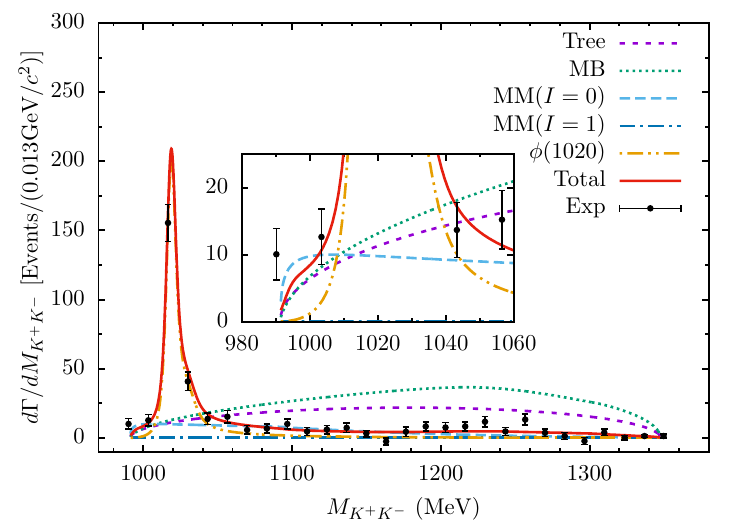}
    \put(16,60){\large (a)}
\end{overpic}
\end{minipage}

\vspace{0.5cm}
\begin{minipage}{1\columnwidth}
\centering
\begin{overpic}[scale=0.65]{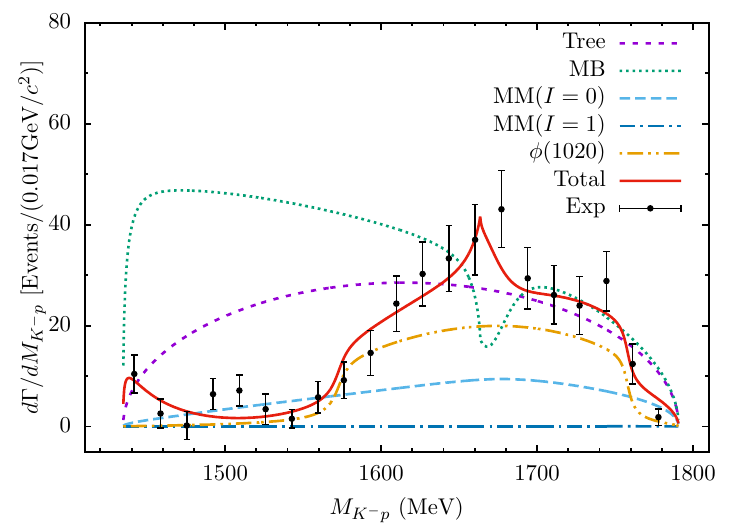}
    \put(15,60){\large (b)}
\end{overpic}
\end{minipage}

\vspace{0.5cm}
\begin{minipage}{1\columnwidth}
\centering
\begin{overpic}[scale=0.65]{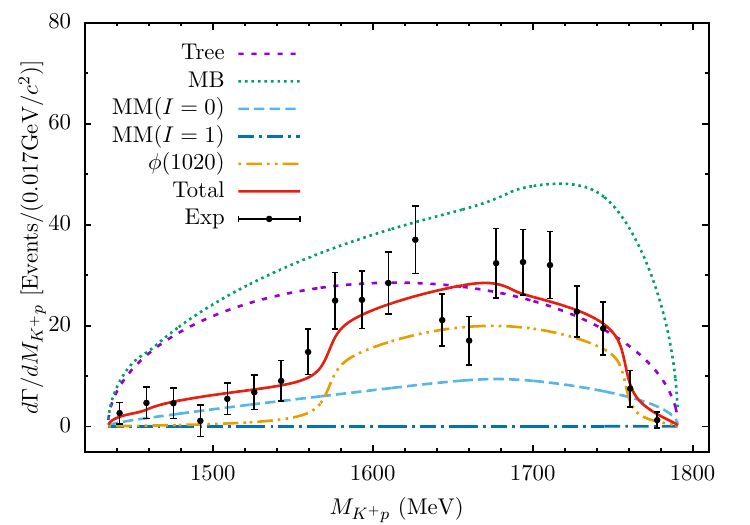}
    \put(85,60){\large (c)}
\end{overpic}
\end{minipage}
\caption{$K^+K^-$ (a), $pK^-$ (b), and $pK^+$ (c) invariant mass distributions of the $\Lambda_c^+ \to pK^+K^-$ decay with the fitted parameters of Model B.}
\label{fig:with_para8}
\end{figure}

\begin{figure}[htbp]
		\centering
		\includegraphics[scale=0.9]{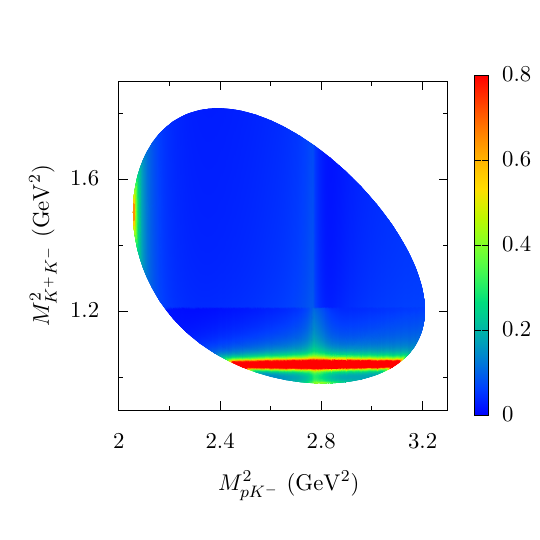}	
 \caption{Dalitz plot of the process $\Lambda_c^+ \to p K^+ K^-$ with Model B.}
\label{fig:Dalitz}
\end{figure}

In model A, we have six free parameters: $V_{\rm MB}$, $V_{\rm MM}$, and $V_{\phi(1020)}$ represent the relative weight factors; $\phi_1$ and $\phi_2$ denote the interference phases; and the parameter $r_T$ describes the ratio of the transverse amplitude to the longitudinal one for the $\phi(1020)$ resonance. We have fitted the free parameters to the mass distributions measured by BESIII~\cite {BESIII:2026dkj}, and tabulated them in Table~\ref{tab:fit_params} (Model A). The fit yields  $\chi^2/{\rm d.o.f.} = 1.28$. With the fitting parameters, we have calculated the invariant mass distributions for the process $\Lambda_c^+ \to p K^+ K^-$, as shown in Fig.~\ref{fig:with_para6}.

The total contribution is shown by the red solid line in Fig.~\ref{fig:with_para6}, and the data and error bars are taken from the experimental measurements of BESIII~\cite{BESIII:2026dkj}. The purple dotted line represents the tree-level contribution; the green dotted line corresponds to the contribution from the meson-baryon interaction; the light blue dashed line depicts the meson-meson interaction in the isospin $I=0$ channel, whereas the dark blue dash-dotted line corresponds to the $I=1$ coupled channel; and the orange dash-dotted line illustrates the contribution from the $\phi(1020)$ resonance.

In the $K^+K^-$ invariant mass distribution shown in Fig.~\ref{fig:with_para6}(a), a prominent peak is observed around $1020~\mathrm{MeV}$, which is attributed to the intermediate $\phi(1020)$ resonance. The pronounced near-threshold enhancement mainly arises from the isoscalar $f_0(980)$ state with isospin $I=0$, and the contribution from the isospin $I=1$ meson-meson interaction is negligible. 

Meanwhile, a clear cusp structure around $1670~\mathrm{MeV}$ is observed in the $pK^-$ invariant mass distribution of Fig.~\ref{fig:with_para6}(b), associated with the resonance $\Lambda(1670)$.  
One can find that the contribution from the meson-baryon final-state interactions gives rise to a dip structure around $1670~\mathrm{MeV}$ in the $M_{pK^-}$ distribution, which is consistent with the characteristic signature of the $\Lambda(1670)$ resonance described in Ref.~\cite{Li:2026umb}. 
Once all partial amplitudes are added coherently, this dip develops into a pronounced cusp, demonstrating that the interference plays a decisive role in shaping the line shape. This observation further indicates that the threshold effect of the $\eta\Lambda$ channel influences the line shape of the $\Lambda(1670)$. Furthermore, near the $pK^-$ threshold, a prominent enhancement is observed, which originates from the resonance $\Lambda(1405)$. 

In contrast, the $M_{pK^+}$ distribution in Fig.~\ref{fig:with_para6}(c) exhibits a considerably smoother line shape. Since the $p K^+$ system has isospin $I = 1$ and strangeness $S = 1$, it does not couple to any known baryon. The broad structure around $1600$--$1700~\mathrm{MeV}$ is dominated by the $I=1$ meson-baryon coupled channel FSI and the tree-level background, with an additional contribution from the kinematic reflection of the $\phi(1020)$ band.

It it notable that in Fig.~\ref{fig:with_para6}(a), one can find that the predicted  $\phi(1020)$ peak is higher than the BESIII data. Furthermore, we performed another fit by taking the mass and width of the $\phi(1020)$ as  free parameters. The fitted parameters are also tabulated in Table~\ref{tab:fit_params} (Model B). The fit yields $\chi^2/{\rm d.o.f.}= 1.06$, which represents an improved fit quality relative to Model A ($\chi^2/{\rm d.o.f.}=1.28$).

We have recalculated the invariant mass distributions using the fitted parameters of Model B and present the results in Fig.~\ref{fig:with_para8}. The fitted mass and width of $\phi(1020)$ are close to the nominal values of RPP~\cite{ParticleDataGroup:2024cfk}. In the $K^+K^-$ invariant mass spectrum, we again observe a prominent $\phi(1020)$ peak as well as a threshold enhancement associated with the $I=0$ $f_0(980)$ state. In the $pK^-$ invariant mass spectrum, a clear enhancement from the $\Lambda(1405)$ state appears near threshold, and a pronounced cusp structure related to the $\Lambda(1670)$ resonance emerges around 1670~MeV. These results demonstrate that our analysis provides a good description of the experimental measurements from BESIII~\cite{BESIII:2026dkj}.

In addition, as shown in Fig.~\ref{fig:Dalitz}, we present the Dalitz plot of $M^2_{pK^-}$ versus $M^2_{K^+K^-}$ for the $\Lambda_c^+ \to p K^+ K^-$ decay using Model B. One can clearly observe the signals of the resonances $\Lambda(1405)$, $\Lambda(1670)$, $f_0(980)$, and $\phi(1020)$.

Finally, using the branching fraction of $\Lambda_c^+\to pK^+K^-$ measured by BESIII~\cite{BESIII:2026dkj}, we calculate the branching fractions of the subprocesses by integrating the differential decay widths. This yields
$\mathcal{B}(\Lambda_c^+\to \Lambda(1405)K^+,\Lambda(1670)K^+\to pK^+K^-)=0.54\times 10^{-3}$,
which is consistent with the experimental value of $(0.36\pm 0.18)\times 10^{-3}$~\cite{BESIII:2026dkj}.
The calculated branching fractions
$\mathcal{B}(\Lambda_c^+\to pf_0(980)\to pK^+K^-)=0.38\times 10^{-3}$ and
$\mathcal{B}(\Lambda_c^+\to p\phi\to pK^+K^-)=0.62\times 10^{-3}$
are also consistent with the BESIII results of $(0.43\pm 0.17)\times 10^{-3}$ and
$(0.62\pm 0.05)\times 10^{-3}$~\cite{BESIII:2026dkj}, respectively.

\section{Summary}
\label{Sec:summary}
Recently, the BESIII Collaboration analyzed the singly Cabibbo-suppressed decay $\Lambda_c^+ \to p K^+ K^-$, performing a partial wave analysis of this charmed baryon decay~\cite{BESIII:2026dkj}, which provides an important laboratory to study the scalar mesons and light baryons.

Motivated by the BESIII results, in this work we have analyzed the process $\Lambda_c^+ \to p K^+ K^-$ by considering the intermediate states $\Lambda(1405)$ and $\Lambda(1670)$  dynamically generated via the $S$-wave pseudoscalar meson-octet baryon interaction, and the $f_0(980)$ and $a_0(980)$ resonances, dynamically generated through the $S$-wave pseudoscalar meson-pseudoscalar meson interaction within the chiral unitary approach. In addition, we have also included the contribution from the vector meson $\phi(1020)$ in the $P$-wave. Since the bin width used in the experimental invariant mass spectrum is  larger than the width of the $\phi(1020)$ in RPP~\cite{ParticleDataGroup:2024cfk}, the theoretical distribution is convolved with the detector mass resolution function.

By performing a fit to the BESIII measurements of the $K^+K^-$, $pK^-$, and $pK^+$ invariant mass distributions, we obtain the best fit  $\chi^2/\mathrm{d.o.f.}=1.06$, indicating an overall satisfactory description of the experimental data. In particular, the dynamically generated $\Lambda(1405)$ and $\Lambda(1670)$ resonances in our model successfully reproduce the threshold enhancement and the cusp structure around $1670~\mathrm{MeV}$, respectively, in the $pK^-$ invariant mass distribution. In the $K^+K^-$ invariant mass distribution, the contribution from the $f_0(980)$ provides a good description of the enhancement near the $K^+K^-$ threshold, and the $a_0(980)$ contribution is negligible. 

Moreover, a pronounced peak associated with the $\phi(1020)$ resonance is observed around $1020~\mathrm{MeV}$. The sizable strength of this feature indicates that the $\phi(1020)$ contribution plays an important role in the $\Lambda_c^+\to pK^+K^-$ decay.

Future measurements of the $\Lambda_c^+ \to pK^+K^-$ decay with improved statistics and invariant mass resolution will be essential for clarifying the underlying decay mechanisms and the interplay among different final-state interactions. Such studies will provide more detailed information on the properties and internal structures of low-lying baryon resonances and light scalar mesons, and will further deepen our understanding of the nonleptonic weak decay dynamics of charmed baryons.

\section{ACKNOWLEDGMENTS}

This work is supported by the National Key R\&D Program of China (Nos. 2024YFE0105200 and 2023YFA1606703), the National Natural Science Foundation of China under Grant Nos. 12475086, 12192263, 12205075, 12575094, 12435007, W2543006, and 12361141819.

\appendix
\section{Derivation of the amplitude for the $\phi(1020)$ contribution}
\label{app:phi}

In this appendix, we provide a detailed derivation of the decay amplitude
$\mathcal{T}_{s_p s_{\Lambda_c}}^{\phi}$ for the cascade decay
\begin{equation}
\Lambda_c^+ \to p\,\phi(1020) \to p\,K^+K^-,
\end{equation}
which enters the total decay amplitude of the $\Lambda_c^+ \to pK^+K^-$
process. The derivation follows the Jacob--Wick helicity formalism~\cite{Jacob:1959at} together with the standard derivative coupling for the
$\phi K^+K^-$ vertex~\cite{Klingl:1996by}.

The $\Lambda_c^+$ baryon has spin $1/2$, as does the final-state proton,
while the $\phi(1020)$ is a spin-1 vector meson. The decay can be
decomposed into the two-step chain
\begin{equation}
\Lambda_c^+ \to p\,\phi(1020), \qquad \phi(1020) \to K^+K^-.
\end{equation}
For the weak decay $\Lambda_c^+ \to p\,\phi$, the helicity of the initial
$\Lambda_c^+$ is $\lambda_{\Lambda_c}=s_{\Lambda_c}=\pm 1/2$, that of the
proton is $\lambda_p=s_p=\pm 1/2$, and that of the $\phi$ is
$\lambda_\phi = -1,\,0,\,+1$. The corresponding helicity amplitudes are
denoted by $H_{\lambda_p,\lambda_\phi}$. Since the two kaons are spinless
pseudoscalars, the helicity of the $\phi$ is conserved in the strong decay
$\phi \to K^+K^-$, whose helicity amplitude is denoted by $C$.

In the helicity formalism, the total amplitude of the cascade decay is the
sum over the intermediate $\phi$ helicity $\lambda_\phi$ of the product of
the two-step amplitudes and the corresponding Wigner $D$ functions,
\begin{equation}
\mathcal{T}_{s_p s_{\Lambda_c}}
= C \sum_{\lambda_\phi=-1,0,+1}
  H_{\lambda_p,\lambda_\phi}\,
  D^{1}_{\lambda_\phi,0}(\Omega_2)\,
  D^{1/2}_{s_{\Lambda_c},s_p}(\Omega_1),
\label{eq:app-A}
\end{equation}
where $\Omega_1$ and $\Omega_2$ denote the Euler angles associated with the
decay $\Lambda_c^+ \to p\,\phi$ and $\phi \to K^+K^-$, respectively.

The angle $\theta$ in the main text is defined as the angle between the
$K^+$ and the proton $p$ in the $K^+K^-$ rest frame. The relevant Wigner
$d$ functions for the spin-1 $\phi$ meson are
\begin{equation}
d^{1}_{0,0}(\theta) = \cos\theta, \,
d^{1}_{+1,0}(\theta) = -\frac{\sin\theta}{\sqrt{2}}, \,
d^{1}_{-1,0}(\theta) = +\frac{\sin\theta}{\sqrt{2}}.
\end{equation}
Substituting these into Eq.~\eqref{eq:app-A}, the sum over
$\lambda_\phi$ reads
\begin{eqnarray}
    \sum_{\lambda_\phi} H_{\lambda_p,\lambda_\phi}\, d^{1}_{\lambda_\phi,0}(\theta)
&=& H_{\lambda_p,0}\cos\theta
  - \frac{H_{\lambda_p,+1}}{\sqrt{2}}\sin\theta \nonumber \\
&&  + \frac{H_{\lambda_p,-1}}{\sqrt{2}}\sin\theta.
\label{eq:app-B}
\end{eqnarray}

The helicity-$\lambda_\phi = 0$ component is spin independent and produces
the $\cos\theta$ angular dependence, while the transverse components with
$\lambda_\phi = \pm 1$ produce the $\sin\theta$ dependence. Because both
the $\Lambda_c^+$ and the proton carry spin $1/2$, the transverse helicity
transfer is equivalent to an $SU(2)$ rotation of the initial $\Lambda_c^+$
spinor, which is encoded in the Pauli matrix term
$\boldsymbol{\sigma}\cdot\hat{\mathbf{n}}$, where $\hat{\mathbf{n}}$ is
the unit vector normal to the decay plane.

Introducing the longitudinal amplitude $\mathcal{C}_0 \equiv
H_{\lambda_p,0}$ and the transverse amplitudes $\mathcal{C}_{\pm} \equiv
H_{\lambda_p,\pm 1}$, and normalizing the overall scale by
$\mathcal{C}_{\phi}(M_{K^+K^-}) \equiv \mathcal{C}_0$, the amplitude can be
reorganized into the compact $2\times 2$ matrix form
\begin{equation}
\mathcal{T}_{s_p s_{\Lambda_c}}^{\phi}
= \mathcal{C}_{\phi}(M_{K^+K^-})\,
  \chi_{s_p}^{\dagger}
  \left[
    \cos\theta\,\mathbf{1}
    + i\,r_T \sin\theta\,\boldsymbol{\sigma}\cdot\hat{\mathbf{n}}
  \right]
  \chi_{s_{\Lambda_c}},
\label{eq:app-22}
\end{equation}
where $\chi_{s_{\Lambda_c}}$ and $\chi_{s_p}$ are the Pauli spinors of the
initial $\Lambda_c^+$ and the final proton, respectively, $\mathbf{1}$ is
the $2\times 2$ identity matrix, and $r_T$ quantifies the transverse amplitude
relative to the longitudinal one. 

For an unpolarized $\Lambda_c^+$ and without measuring the proton
polarization, the spin-averaged squared amplitude reads
\begin{eqnarray}
\left|\mathcal{T}^{\phi(1020)}\right|^2
&=& \frac{1}{2}\sum_{s_{\Lambda_c},s_p}
  \left|\mathcal{T}_{s_p s_{\Lambda_c}}^{\phi}\right|^2 \nonumber \\
&=& \left|\mathcal{C}_{\phi}(M_{K^+K^-})\right|^2
  \left[\cos^2\theta + |r_T|^2\sin^2\theta\right].
\label{eq:app-25}    
\end{eqnarray}

The cross term proportional to
$2\cos\theta\sin\theta\,\mathrm{Tr}(\boldsymbol{\sigma}\cdot\hat{\mathbf{n}})$
vanishes after the spin summation, since
\begin{equation}
\mathrm{Tr}(\boldsymbol{\sigma}\cdot\hat{\mathbf{n}}) = 0.
\label{eq:app-26}
\end{equation}

The transverse parameter $r_T$ can be related to the helicity amplitudes of
the weak decay $\Lambda_c^+ \to p\,\phi$ through
\begin{equation}
|r_T|^2
= \frac{|H_{+\frac{1}{2},+1}|^2 + |H_{-\frac{1}{2},-1}|^2}
       {2\left(|H_{+\frac{1}{2},0}|^2 + |H_{-\frac{1}{2},0}|^2\right)},
\label{eq:app-27}
\end{equation}
where the numerator collects the transverse ($|\lambda_\phi| = 1$) helicity
amplitudes and the denominator collects the longitudinal
($\lambda_\phi = 0$) ones, the factor $2$ arising from the spin average.
Because the helicity amplitudes $H_{\lambda_p,\lambda_\phi}$ are determined
by the weak production dynamics, $r_T$ is not fixed by the quantum numbers
of the $\phi(1020)$ alone and is treated as a free parameter in the present
analysis.

This fixes the momentum dependence of the overall coefficient
$\mathcal{C}_{\phi}$ in Eq.~\eqref{eq:app-22}:
\begin{equation}
\mathcal{C}_{\phi}(M_{K^+K^-})
= \frac{V_{\phi(1020)}\,
        |p_{K^+}^{*}|\,|p_{p}^{*}|}
       {D_{\phi}(M_{K^+K^-})},
\label{eq:app-23}
\end{equation}
where $D_{\phi}(M_{K^+K^-})$ is the Breit--Wigner denominator given in
Eq.~(\ref{eq:phi_bw_denominator}) of the main text, and $|p_{K^+}^{*}|$ and $|p_{p}^{*}|$ are the
momenta of the $K^+$ and the proton in the $K^+K^-$ rest frame. The overall coupling
$V_{\phi(1020)}$, determined by fitting to the experimental data, absorbs
the weak production coupling, the $\phi K^+K^-$ coupling, and
convention-dependent normalization factors.



\end{document}